\documentclass[12pt,pdflatex]{article}
\usepackage{tabularx}
\usepackage{graphicx}
\usepackage{subfigure}
\usepackage{dcolumn}
\usepackage{hyperref}
\usepackage[T1]{fontenc}
\usepackage[utf8]{inputenc} %
\usepackage[english]{babel}
\usepackage{amsmath,bm}
\usepackage{amstext,amsthm,amscd}
\usepackage{amsfonts,txfonts}
\usepackage{amssymb}
\usepackage{multirow, array}
\usepackage{mathrsfs}
\usepackage[affil-it]{authblk}
\usepackage[english]{babel}
\usepackage[left=2cm,right=2cm,top=2cm,bottom=2cm]{geometry}
\usepackage[usenames]{color}

\definecolor{MHcolor}{rgb}{.5,.5, 1}

\usepackage{epstopdf}
\begin{document}

\title{Top quark production through flavor violating neutral currents within the 2HDM type-III }

\author[1]{M. Gómez-Bock\footnote{melina.gomez@upaep.mx}}

\author[2]{W. Gonzalez-Olivares \footnote{w\_gonzalez\_olivares@tec.mx} }

\author[3]{M. Hentschinski \footnote{martin.hentschinski@udlap.mx}}

\author[4]{A. Rosado \footnote{rosado@ifuap.buap.mx}}

\author[5]{S. Rosado-Navarro \footnote{sebastian.rosado@protonmail.com}}

\affil[1]{\small Area of Mathematics, Universidad Popular Aut\'onoma del Estado de Puebla, 21 sur 1103, Barrio de Santiago. C.P. 72410, Puebla, Pue. M\'exico.}

\affil[2]{School of Science and Engineering, Tecnol\'ogico de Monterrey-Puebla. Vía Atlixcáyotl 5718, Reserva Territorial Atlixcáyotl, C.P.72453, Puebla, Pue. M\'exico. }

\affil[3]{Dpt. of Actuarial Science, Physics and Mathematics, Universidad de las Américas Puebla, Ex. hda. Sta. Catarina m\'artir s/n San Andr\'es Cholula, C.P.72810, Pue. M\'exico. }

\affil[4]{Physics Institute,  Benem\'erita Universidad Aut\'onoma de Puebla, Av. San Claudio y Blvd. 18 Sur, Col. San Manuel, Ciudad Universitaria, C.P. 72570, Puebla, Pue. M\'exico. }

\affil[5]{Centro Interdisciplinario de Investigación y Ense\~nanza de la Ciencia, Benemérita Universidad Autónoma de Puebla, Av. San Claudio y Prol. 24 sur, Col. San Manuel Ciudad Universitaria,
C.P. 72570, Puebla, Pue. México. }

%

\maketitle

\begin{abstract}
We explore the parameter space for processes with fermionic flavor violation within the 2HDM type III, processes of Flavor Changing Neutral Currents are present at LO as dynamics beyond the Standard Model are included. We analyze the possible Yukawa couplings beyond the SM involving the quark top, as the experimental signatures would be a clear signal to conclude about the model. The processes we analyzed are same sign top pair production at the $pp$ collisions,  and  single top quark production in $e p$ collisions, through DIS.
To examine the scenarios that could exhibit such exotic events, we perform a scan in the parameter space to determine the observation possibilities for FCNC. We show the results considering the model parameters, coming from modified Yukawa interactions, $\vert\tilde{\chi}_{ij}^{f}\vert$ of the order one. In addition we explore the order of magnitude of possible single top production via $\mu p$, at the proposed muon-proton collider. 
We found scattering values for these exotic processes of  up to $\mathcal{O}(10^{-2})~pb$.
\end{abstract}

                      

\maketitle

\section{\label{Intro}Introduction}
The original phenomenological structure of the Standard Model of Elementary Particle Physics  is based on the use of conserved and discrete quantum numbers which determine the allowed dynamical processes.  They exhibit a set of restricted Flavor Changing Charged Currents (FCCCs) and suppressed Flavor Changing Neutral Currents (FCNCs).  
This structure helps to identify possible processes to be observed experimentally and reduces furthermore the structure of the model.
Nevertheless, {\it Flavour Physics} has shown a much more richer phenomenology than  the one predicted by the Standard Model (SM). A major breakthrough is the evidence for neutrino mixing  \cite{heeger2005evidence}, also possibly non-universality of weak leptonic meson decay \cite{falkowski2015lepton} or the {\it B-anomalies}\cite{LHCb:2021trn}. Moreover, the recent results on {\it (g-2)} \cite{Muong-2:2023cdq} and the CDF analysis on $W$ boson mass\cite{CDF:2022hxs} may suggest beyond the SM weak structure.  These observations could in principle affect the original flavor structure of the SM, which has been constructed as an effective theory from the available low energy experimental data. They therefore seem to indicate the limits of the SM 
 and hint at the need to extend the flavor structure of the SM.
 
In order to know with certainty if we are in extended model domains,  an excellent candidate to study such processes is  given by events which are associated with the production of a top quark, since its mass  is  the largest of the known elementary particles and easy to track down as it would not hadronize.
The production of a pair of top or anti-top quarks at the LHC \cite{single}, allows to constrain a potential transition of a charm quark into a top quark via a t-channel exchange of a flavor violating Higgs boson. From experimental point of view, the exclusive same sign double top production, with no other final states, has so far been searched for by  both the CMS \cite{CMS:2011gff} and  the ATLAS \cite{ATLAS:2012iws} collaborations.
Furthermore, single top production at an electron-proton collider would on the other hand, serve to  constrain the coupling of a flavor violating Higgs to the leptonic sector, which is hard to control at an hadron-hadron collider.  A first bound was already established at the HERA experiments, where $\sigma(ep\to etX)< 0.25pb$  for an energy of$\sqrt{s}=319$~GeV,  \cite{H1:2009yuy}. We could expect an improvement of this bound at an future LHeC \cite{Gao:2021plf}. 
A muon-proton collider, as proposed in \cite{Acosta:2022ejc},  would on the other hand give access to the muon-to-tau transition through $t$-channel exchange of a flavor violating Higgs boson and in this way further constrain the content of the model. \\

Within the Standard Model (SM), quarks acquire their mass through
spontaneous symmetry breaking \cite{Wells:2009kq}, while the mechanism cannot explain
the mass spectrum and the resulting mass hierarchy.  From the theoretical structure of the Higgs mechanism, the coefficients are directly related to the mass values of the particles involved nevertheless, the Yukawa couplings are only determined experimentally, see \cite{Sirunyan:2018lzm} for a recent analysis. In order to account for this enriched flavor structure and to explain it, it is needed to  extend the SM.  
A minimal extension of the SM which allows for flavor changing neutral currents at the level of the Lagrangian is provided by the Two Doublet Higgs Model with Flavour Violation (FV) at Leading Order (LO), the 2HDM-III, which we will discuss in more detail in the next section. \\ 

Top quark involved processes within the 2HDM-II and III models have been explored in literature. A general structure within THDM with minimal flavor violation in the context of an {\it Effective Lagrangian}, using a structure in terms of Wilson coefficients, which are bounded with meson process is presented in \cite{Crivellin:2012ye,Crivellin:2015hha}. Work around same sign top production, within a parametrization of the Yukawa matrices, separating the off-diagonal interactions and exploring single and triple top production, using elements on these matrices as free parameters, they calculate processes through heavy scalar Higgs bosons with up- and charm -quarks  \cite{Hou:2018zmg,Hou:2020ciy, Hou:2020chc}.  In particular, considering two Yukawas matrices accounting for the interaction on both scalar Higgs doublets with both types of quarks in the interaction basis is done in \cite{Diaz-Cruz:2004wsi}, giving a complete analytical structure and digonalization of the fermionic mass matrix, the matrix elements have to accomplish for mass relations and hermitian restrictions. In \cite{Arhrib:2016tni}, based on the same last parametrization structure of the Yukawa matrices, flavor violation process are considered in order to bound the parameter space for 2HDM-III; from which we use the bound for flavor violation top process  $h\to c t$, given in terms of the parameters of the model. This has been established as $BR(t\to c, h(H))<10^{-3}$.

The paper is organized as follows. In section 2, we present the  FCNC stemming from Yukawa interactions within the 2HDM-III.
In section 3, we calculate the cross sections for  double top quark production at the LHC and single top quark production at $ep$ collider, via Deep Inelastic Scattering (DIS),  through FCNC in the context of the THDM-III, using the Parton Model with five scheme PDFs. In section 4, we perform numerical calculation for the processes giving constraints on the parameters of the model due to experimental bounds. Finally, we summarize our conclusions.


\section{Flavor violating processes within the 2HDM-III model}
Flavor violating processes have been widely studied in the literature, in particular for the so called exotic decays  via a neutral Higgs boson, using an effective Lagrangian framework \cite{Goldouzian:2014nha}, through higher dimension operators in the context of THDM \cite{Buschmann:2016uzg}, and directly from the Lagrangian in the specific scenario of THDM-III  \cite{Das:2015kea, Arhrib:2016tni}.
Extensions on the Higgs mechanism with more than one Higgs doublet are still widely explored in the literature, {\it e.g.} \cite{Bento:2017eti}.
Adding another Higgs doublet to the Standard Model is the simplest extension possible of the SM Higgs sector. It leads to the Two Higgs Doublets Models (THDM), and is also the lowest possible structure for the scalar sector  in the context of Supersymmetry. THDM models differ among each other,
based on a discrete symmetry added to the model in order to avoid FCNC, as the masses of the fermions originate from two doublets in contrast to one doublet in the Standard Model. To establish the precise model one uses direct search of new scalar particles, but also  indirect searches such as rare processes of flavor violation. \\

The actual possibility of having FCNC has been discussed in the literature. The initially studied versions of the Higgs doublet model completely excluded this possibility as in 2HDM type I and II \cite{arhrib2016two}, the specific 2HDM with FV tree level structure was studied in \cite{Diaz-Cruz:2004wsi,melina}, in the so-called THDM-III. In this article we claim that while those flavor mixing couplings must be small, as is reported in other processes \cite{das2016flavor}, they do not need to be zero, as in the SM.\\

In the following we review the relevant elements of the 2HDM-III model for our study. 
The 2HDM-III Lagrangian of the Yukawa sector has the form
\begin{equation}
L_{Y}=\sum_{a,i}Y_{a}^{i} \overline{F}_{L}^{i}\Phi_a
f_{R}^{i}+h.c.,
\end{equation}
where $F_{L}$ denotes the fermion doublet (left-handed), $f_{R}$ is the fermion singlet (right-handed), and $\Phi_a$ are the Higgs doublets $(a=1,2)$. Considering three generations, the coefficient $Y_{a}^{i}$ can be expressed as a $3\times3$ matrix and tagged as $Y_{a}^{l}, Y_{a}^{u}, Y_{a}^{d}$, for leptons, $u$ and $d$ type quarks respectively, while we take neutrinos to be massless.  
For type III models, the two Higgs doublets  couple to  two quarks with different isopin \cite{DiazCruz:2004tr}.  This is a particularly interesting  feature of this model, since it allows to explain in a more natural way the mass hierarchy of the Standard Model fermions.
In the 2HDM-III we have three types of free parameters that must be determined from experiment, \emph{i.e.} scalar masses, Yukawa couplings and the ratio of vacuum expectation values. Then we have the following Yukawa couplings to fermions:

\begin{equation}
\begin{split}
L_{Y}^{q} &=Y_{1}^{u}\overline{Q}^{\prime}_{L}\tilde{\Phi}_1 u_{R}^{\prime}+Y_{2}^{u}\overline{Q}_{L}^{\prime}\tilde{\Phi}_2 u_{R}^{\prime}+Y_{1}^{d}\overline{Q}_{L}^{\prime}\Phi_1 d_{R}^{\prime}+Y_{2}^{d}\overline{Q}_{L}^{\prime}\Phi_2 d_{R}^{\prime}+h.c.,\\
\end{split}
\end{equation}
where $\tilde{\Phi}_{1,2}=i\sigma_2\Phi^{*}_{1,2}$  and $\sigma_2$ is the Pauli matrix. The charged leptonic sector has a   similar form to the one of
the $d-type$ quark, and is obtained from the latter by replacing
$d_{i}\rightarrow\,l_{i}$, including the masses.
After spontaneous symmetry breaking, each of the two doublets acquire vacuum expectation values (vevs),  $v_{1,2}$. Expanding the scalar complex fields over this vevs, one obtains:
\begin{align}
\Phi_1&=\begin{pmatrix}
\phi^{+}_{1} \\
\frac{v_1}{\sqrt{2}}+\phi^{0}_{1}
\end{pmatrix}
; &
\Phi_2&=\begin{pmatrix}
\phi^{+}_{2} \\
\frac{v_2}{\sqrt{2}}+\phi^{0}_{2}
\end{pmatrix},
\end{align}
which yields the following form of the mass matrices:\\

\begin{equation}
M_f=\frac{1}{\sqrt{2}}(v_{1}Y_{1}^{f}+v_{2}.
Y_{2}^{f}),\hspace{0.5cm} f=u,d,l.
\end{equation}
In the physical basis, $M_f$ is diagonal but not necessary are each of the two Yukawa matrices. In order to diagonalize  analytically, we reduce the possible $3\times3$ flavor fermion mass matrices by a proposed {\it ansatz} with a hierarchical structure, which is  based on a textures form (zero for some flavor mixing elements guided by experimental data), as was first studied in \cite{antaramian1992flavor,sher1991rare}. For the two hermitian
Yukawa matrices, we  consider a more general flavor structure with a 4-texture, considering both types of quarks. Following \cite{DiazCruz:2004tr}, we have 
\begin{eqnarray}
M_f=
\begin{pmatrix}
0 & C_l &0 \\
C^*_l & \tilde{B}_l & B_l \\
0 & B^*_l & A_l \\
\end{pmatrix} ;\; & 
Y_1=
\begin{pmatrix}
0 & C_1 &0 \\
C^*_1 & \tilde{B}_1 & B_1 \\
0 & B^*_1 & A_1 \\
\end{pmatrix} ; \;
Y_2=
\begin{pmatrix}
0 & C_2 &0 \\
C^*_2 & \tilde{B}_2 & B_2 \\
0 & B^*_2 & A_2 \\
\end{pmatrix},\notag \\
\label{textures}
\end{eqnarray}
where the elements of the matrix should be taken with a hierarchy imposed by the measured fermion masses,  $|A_l| \gg |\tilde{B}_l|, |B_l|, |C_l|$, which
can be diagonalized through
\begin{eqnarray}
 \bar{M}^{diag}_{u}=V^{u}_{L}M_{u}V^{u \dagger}_{R},\notag \\
 \bar{M}^{diag}_{d}=V^{d}_{L}M_{d}V^{d \dagger}_{R},\notag \\
 \bar{M}^{diag}_{l}=O^{l}_{L}M_{l}O^{l \dagger}_{R}.
\end{eqnarray}
Here, as usual, $V_{CKM}=V^{u}_{L}V^{d\dagger}_{L}$, and 
\begin{eqnarray}
\tilde{Y}^{q}_{1,2}=V^{q}_{L}Y^{q}_{1,2}V^{q \dagger}_{R}\; \text { and }\;  &
\tilde{Y}^{l}_{1,2}=O^{l}_{L}Y^{l}_{1,2}O^{l \dagger}_{R},
\end{eqnarray}
yields the CKM matrix 
and $q=u,b$.
We further note the following relation between the two Yukawa matrices for each fermion type:
\begin{eqnarray}
\displaystyle
\tilde{Y}_{1}^{d}=
\frac{\sqrt{2}}{v\cos\beta}\bar{M}_{d}-\tan\beta\tilde{Y}_{2}^{d} \\
\tilde{Y}_{1}^{l}=\frac{\sqrt{2}}{v\cos\beta}\bar{M}_{l}-\tan\beta\tilde{Y}_{2}^{l} \\
\tilde{Y}_{2}^{u}=\frac{\sqrt{2}}{v\sin\beta}\bar{M}_{u}-\cot\beta\tilde{Y}_{1}^{u} & &
\label{rotyukawas}
\end{eqnarray}

Having an extra Higgs  scalar doublet in the THDM, 
as it is known \cite{Gunion:1984yn,Gomez-Bock:2007azi}, $\alpha$ is the rotation angle for CP-even physical neutral Higgs bosons $h^0$ and $H^0$ states, and $\beta$ is the angle associated with the Goldstone states basis,  $\tan \beta= v_2/v_1$. In the physical basis, omitting Goldstone contributions, we have for the fermions couplings with neutral scalars:
\begin{eqnarray}
{\cal{L}}_Y^{q} & = &\frac{g}{2}
\left\{ \bar{u}_{i}
\left[\left(\frac{m_{u_{i}}}{m_W}\right)\frac{\cos\alpha}{\sin\beta} \delta_{ij}- \frac{\sqrt{2} \,
\cos(\alpha - \beta)}{g \, \sin\beta}
(\tilde{Y}_1^u)_{ij}\right]u_{j}h^{0} \right.
\nonumber \\
&+& \left.\bar{d}_{i}
\left[-\left(\frac{m_{d_{i}}}{m_W}\right)\frac{\sin\alpha}{\cos\beta} \delta_{ij}+ \frac{\sqrt{2} \,
\cos(\alpha - \beta)}{g \, \cos\beta}(\tilde{Y}_2^d)_{ij}\right]d_{j}
h^{0}\right. \nonumber \\
& & \left. + \bar{u}_{i}\left[\left(\frac{m_{u{i}}}{m_W}\right)\frac{ \, \sin\alpha}{\sin\beta}\delta_{ij}-\frac{\sqrt{2} \, \sin(\alpha - \beta)}{g \, \sin\beta}(\tilde{Y}_1^u)_{ij}\right]u_{j}H^{0}
\right.
\nonumber \\
&+& \left.\bar{d_{i}}\left[\left(\frac{m_{d_{i}}}{m_W}\right)\frac{ \, \cos\alpha}{\cos\beta}\delta_{ij}+
\frac{\sqrt{2} \, \sin(\alpha - \beta)}{g \, \cos\beta}(\tilde{Y}_2^d)_{ij}\right]d_{j}H^{0}\right. \nonumber \\
& &\left. 
+i\bar{u}_{i}\left[-\left(\frac{m_{u_{i}}}{m_W}\right)\cot\beta \delta_{ij} + \frac{\sqrt{2} }{g \, \sin\beta}
(\tilde{Y}_1^u)_{ij}\right]\gamma^{5}u_{j} A^{0}\right. 
\nonumber \\
&+& \left.i\bar{d}_{i}\left[-\left(\frac{m_{d_{i}}}{m_W}\right)\tan\beta \delta_{ij}+  \frac{\sqrt{2} }{g \, \cos\beta}(\tilde{Y}_2^d)_{ij}\right]
\gamma^{5}d_{j} A^{0} \right\}.
\label{lageigenstates}
\end{eqnarray}

The leptonic part is obtained by replacing
$d_{i}\rightarrow\,l_{i}$. We see that for $h^0$ to be the SM-like Higgs (no flavor violation at LO), one needs to set  $\alpha -\beta = \pi /2$ as the decoupling limit \cite{Gunion_2003}.

Using the Cheng-Sher ansatz to reproduce
the mass hierarchy \cite{cheng1987mass}, the Yukawa couplings from the previous Lagrangian
can be described in terms of dimensionless experimental parameters
$\tilde{\chi}_{ij}$ which could have a complex phase, as the matrices in Eq.(\ref{textures}) are Hermitian. In particular negative  $\tilde{\chi}_{ij}$ is possible. Then, the Yukawa matrix elements are 
\begin{equation}
\label{eq:1}
\begin{split}
\left(\tilde{Y}_{2}^{d,l}\right)_{ij} &=\frac{\sqrt{m_{i}^{d,l}m_{j}^{d,l}}}{v}\tilde{\chi}_{ij}^{d,l}\\
\left(\tilde{Y}_{1}^{u,\nu_l}\right)_{ij} &=\frac{\sqrt{m_{i}^{u,\nu_l}m_{j}^{u,\nu_l}}}{v}\tilde{\chi}_{ij}^{u,\nu_l}.
\end{split}
\end{equation}
Similar to  Eq.~\eqref{eq:1},  a large number of proposals to achieve specific fermion mass matrices are possible,  see for instance reference~\cite{3}. The four zero texture matrix fits however quite well with the quark mixing data. It is worth to point out that, as a consequence of regarding $\tilde{\chi}_{ij}^{f}$ as experimental parameters, we will be  able to define a range where it would be feasible to measure this  FCNC processes. 
For type III models  the two doublets are furthermore coupled to the two types of quarks (up and down) \cite{DiazCruz:2004tr}.  These features of the model would explain in a more natural way the mass hierarchy of the SM fermions.\\

\section{Flavor violating top production within the 2HDM-III model}

As explained in the previous section, within the 2HDM-III model the Lagrangian Eq. (\ref{lageigenstates}) yields  couplings between quarks and leptons of different families, but  with identical isospin. In this work we search for small but non zero flavor changing neutral currents (FCNC) at tree level, which are caused by the extended Higgs sector. In the following we explore experimental signals of such non-zero couplings, assuming that the light Higgs boson of the model $h^0$ is the Higgs boson found at LHC with mass $m_h = 125$~GeV/$c^2$\cite{ATLAS:2012yve,CMS:2012qbp}. 
In the following we will explore this possibility, for both FV transitions between quarks, as well as for FV in leptons. While the former are naturally explored at the Large Hadron Collider, the latter can be accessed in lepton-proton  Deep Inelastic Scattering ($lp$ DIS). In both cases one makes use of the limitation in phase space for the production of two top quarks (hadron-hadron collisions) and a single top quark (hadron-lepton collisions), which leads to a strong suppression of such signals within the Standard Model.

Since the center of mass energy of the LHC is not high enough to allow for top production through QCD evolution, 
the top quark cannot appear in the initial state of a partonic process at the LHC. It must therefore be produced via  weak interactions through a process  -- which involves the exchange of W bosons --  or via strong interaction  -- which requires the production of two top-antitop pairs. In both cases the cross-section is strongly suppressed. For weak interactions the exchange of two $W$ bosons will have the corresponding weak coupling factors reduction. In the strong interactions, it would be due to the limitations of phase space. For a numerical study of Standard Model process using CalcHEP, we found that the dominant sub-process for the SM cross-section for two $t\bar{t}$ pairs production in  proton-proton collision, as obtained from gluon fusion of the order of $0.146 fb$, the up-quark contribution is one order of magnitude lower, whereas the charm contribution is down by three orders of magnitude with respect to gluon-gluon interaction. 

A similar observation holds for the case of lepton-hadron collisions where the center-of-mass energies are in general significantly lower than in hadron-hadron collisions. For a collider setup this in general  due to a significantly lower energy of the leptonic beam in comparison to the hadron beam, which leads to strong reduction in the center-of-mass energy. As a consequence, production of a top-antitop pair is not possible or at least strongly suppressed in such reactions, see Tab.~\ref{SMsigma}
For the LHeC, single top quark production via charged-current DIS is dominant in all the top quark production channels \cite{Gao:2021plf}. Here we use CalHEP \cite{Belyaev:2012qa} to calculate the possible top production within the SM as background, see table \ref{SMsigma}
\begin{table}[ph]
\begin{center}
\caption{Total cross section at LHeC ($\sqrt{s}=1.296~TeV$) for SM background processes regarding the tree level single top production through t-channel. The jets include the top quarks.}
\begin{tabular}{@{}ccc@{}} 
\textrm{process}  & 
\multicolumn{1}{c}{$\sigma (pb)$} &
\textrm{uncertainty} ($\%$)\\
\hline
$e p\to e+1jet$  & 42.6 & 0.00846\\
\hline
$e p\to e+2jet$  & 8.876 & 0.0566  \\
\hline
$e p\to e+2jet+t$  & 0.1702 & 0.8305 \\
\hline
$e p\to \nu_e+W^-+1jet ~\to \nu_e + \mu+\bar{\nu}_{\mu} + 1jet $  & 2.87 & 0.03053 \\
\hline
\end{tabular}
\label{SMsigma}
\end{center}
\end{table}

\begin{figure}[t]
\centering
\includegraphics[width=.4\textwidth]{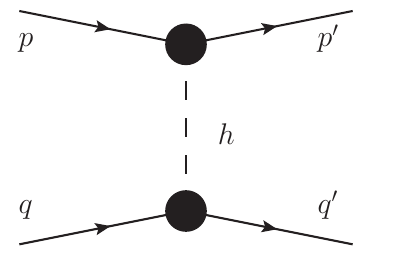}
\caption{Flavor violating process transmitted through a $t$-channel exchange of a neutral Higgs boson of an extended Higgs sector. The final state fermions are either two top quarks (hadron-hadron collisions) or a lepton in combination with a top quark (lepton-hadron collisions).}
\label{fig:mesh1}
\end{figure}

This limitations of phase space is on the other hand absent within the 2HDM-III model, which provides a  direct coupling of up and charm quark to the top  quark through the extended Higgs sector. The generic process is depicted in Fig.~\ref{fig:mesh1}. 
For the Large Hadron Collider, incoming fermions are both up and charm quarks and anti-quarks, where contributions due up (anti-) quarks are strongly suppressed in comparison to the charm contribution
due to the quark mass, which appear in the Higgs couplings  Eq.~\eqref{lageigenstates},  with the suppression factor of the order of  $2.4 \times 10^{-4}$, even if the enhancement due to parton distribution functions is taken into account.  We therefore focus on the contribution due to an initial charm quarks, in absence of partonic top quark and due to  conservation of isospin in our model.
 We will therefore study for hadron-hadron colliders the processes
\begin{align}
    c(p) + c(q) & \to t(p') + t(q') ,\notag \\
    \bar{c}(p) + \bar{c}(q) & \to \bar{t}(p') + \bar{t}(q') ,
\end{align}
which allow to constrain the flavor violating coupling between charm and to quarks in the Higgs sector. For lepton-hadron reactions, one of the incoming quark is replaced by a lepton, {\it i.e.} an electron or a muon. We therefore consider in that case
\begin{align}
    l(p) + c(q) & \to l'(p') + t(q') ,\notag \\
    \bar{l}(p) + \bar{c}(q) & \to \bar{l}'(p') + \bar{t}(q') ,
\end{align}
where $l$  is either an electron or a muon and and $l'$ an electron, a muon, or a tau. As far as past and future collider projects are concerned, such a reaction could be observed at the HERA collider ($\sqrt{s} =314$~GeV) as well as at the LHeC ($\sqrt{s} =1.3$~TeV) and at the proposed muon-ion collider (muon-IC)\cite{acosta2022muon}. The center-of-mass energy of the future Electron Ion Collider ($\sqrt{s} =20-140$~GeV) is on the other hand too low for the production of a top quark, and the observation the proposed process will not be possible. From the experimental side,  exclusive same sign double top production, with no other final states, has so far been searched for by  both the CMS \cite{CMS:2011gff} and  the ATLAS \cite{ATLAS:2012iws} collaborations, where the ATLAS collaboration establishes  $\sigma(tt) < 1.7$~pb  as an upper limit on this production cross-section with 95\% confidence level at $\sqrt{s}=7\,TeV$. \\

Since the process does not involve exchange of color, the color averaged scattering amplitude is identical for quark-quark and quark-lepton scattering and reads:
\begin{align}
\vert\mathscr{M}\vert^{2}(ab \to a'b')&=\dfrac{g^{4}}{64}C_{aa'}(\alpha, \beta) C_{bb'}(\alpha, \beta) 
\cdot 
\frac{m_{a}m_{b}m_{a'}m_{b'}}{m_{W}^{4}} \notag \\
&\times
\frac{\left[t-(m_{a} -m_{a'})^{2}\right]\left[t-(m_{b}-m_{b'})^{2}\right]}{(t-m_{h}^{2})^{2}},
\label{Mampli}
\end{align}
 where quark and lepton masses are neglected against the top mass, whenever both are summed up. We defined the flavor violation couplings for leptonic (or b-type quarks) and t-type quark, respectively as:
\begin{align}
        C_{aa'}(\alpha, \beta)  & = \frac{\cos^2(\alpha - \beta)}{\cos^2(\beta)} |\tilde{\chi}_{ll'}|^{2}, 
    &
     C_{bb'}(\alpha, \beta)  & = \frac{\cos^2(\alpha - \beta)}{\sin^2(\beta)}|\tilde{\chi}_{qq'}|^{2}.
     \label{Cs}
\end{align}
For lepton-hadron process $C_{aa'}$ will come from the leptonic coupling, and $C_{bb'}$ from the t-type quark coupling, while for hadron-hadron collider both will be t-type quark-Higgs coupling.\\

Note that $q\neq q'$ always; however, we could have $l=l'$ in the case of the muon-proton collider. In such case we should also consider the SM coupling as is given in Eq. (\ref{lageigenstates}). Then, for the process $\mu c \to \mu t$, we have:
\begin{align}
\vert\mathscr{M}\vert^{2}(\mu c \to \mu t)&=\dfrac{g^{4}}{64}C_{\mu\mu}(\alpha, \beta) C_{32}(\alpha, \beta) 
\frac{m_{\mu}^{2}m_{c}m_{t}}{m_{W}^{4}}
\frac{t\left[t-(m_{c}-m_{t})^{2}\right]}{(t-m_{h}^{2})^{2}},
\label{Mampli2}
\end{align}
with
\begin{align}
        C_{\mu\mu} (\alpha, \beta) & =\left|-\frac{\sqrt{2}\sin\alpha}{\cos\beta}+ \frac{\cos(\alpha - \beta)}{\cos(\beta)} \tilde{\chi}_{22}^{l}\right|^{2}, 
    \notag\\
     C_{32}(\alpha, \beta)  & = \left|\frac{\cos(\alpha - \beta)}{\sin\beta}\tilde{\chi}_{32}^{u}\right|^{2}
     \label{Cs3}
\end{align}

Before we continue, we must note equation(\ref{Mampli2}) could be zero for certain values of the parameters of the model, since $\tilde{\chi}_{22}^{l}$ is complex, and its complex phase could change the sign, which in turn will imply that the whole process is not viable at LO for 2HDM-III. In this previous case, the relation within parameters of the model takes the following form:
\begin{equation}
 -\frac{\sqrt{2}\sin\alpha}{\cos\beta}+ \frac{\cos(\alpha - \beta)}{\cos\beta} \tilde{\chi}_{22}^{l}=0;  
\end{equation}
taking the extreme complex values for $\tilde{\chi}_{22}^{l} \sim\mathcal{O}(1)$, {\it i.e.} $\chi_{22}^{l}=\pm 1$, which will hold, respectively, the restrictions for $\alpha$ and $\beta$ angles as follow:

\begin{align}
    \sqrt{2}\sin\alpha=\pm \cos(\alpha - \beta) & \Rightarrow 
    \tan\alpha = \frac{\cos\beta}{\mp\sqrt{2}-\sin\beta},
     \label{b-arelation}
\end{align}
which in turns would reflect the possibility of having tree level flavor violation coupling, nevertheless, the combination of mixing angles values given for neutral Higgs sector, will suppress the flavor violation scattering amplitude process.
Having discussed the structure of the couplings, we are able to construct now the analytical hadronic cross-sections, which read
\begin{align}
    \sum_{f=t,\bar{t}}\sigma(pl \to fl')& = \frac{1}{16 \pi} \int_{x_{min}}^1  dx \,   \int_{t_-}^{t^+} dt \,  \left[\frac{|\mathscr{M}|^2 (lc \to l't) }{(xs)^2} \cdot f_c(x, \mu_F)  + \frac{|\mathscr{M}|^2 (l\bar{c} \to l'\bar{t})}{(xs)^2} \cdot f_{\bar{c}}(x, \mu_F)  \right]
    \label{sigmapl}
\end{align}

 for the case of lepton-hadron scattering the center-of-mass energy squared ven as $\hat{s}=xs$, with $x_{min} = \frac{(m_{l'} + m_t)^2}{s}\approx \frac{m_t^2}{s} $,  and 
\begin{align}
    \label{eq:Xsec:hh}
  \sum_{f=t,\bar{t}}    \sigma(pp \to ff) & = \frac{1}{16 \pi} \int_{\frac{4 m_t^2}{s}}^1  dy \,  \cdot \int_{t_-}^{t^+} dt \, \left[ \frac{ |\mathscr{M}|^2 (cc \to tt) }{(ys)^2} {\cal L}_{cc}(y, \mu_F) + \frac{ |\mathscr{M}|^2 (\bar{c}\bar{c} \to \bar{t}\bar{t}) }{(ys)^2} {\cal L}_{\bar{c}\bar{c}}(y, \mu_F)  \right]
\end{align}
for hadron-hadron scattering, where
\begin{equation}
t_{\pm}{(ab;a'b')}=m_a^2+m_{a'}^2-\frac{1}{2\hat{s}}\left[ (\hat{s}+m_a^2-m_{b}^2)(\hat{s}+m_{a'}^2-m_{b'}^2)\mp \lambda^{1/2}(\hat{s},m_a^2,m_{b}^2)\lambda^{1/2}(\hat{s},m_{a '}^2,m_{b '}^2)\right],
\end{equation} 
where
$\lambda(x,y,z)=(x-y-z)^2-4yz$,  and partonic center-of-mass energy squared $\hat{s}=ys$  with
\begin{align}
        {\cal L}_{ab}(y, \mu_F) & = \int_y^1 \frac{dx}{x} f_a(x, \mu_F)f_b(y/x, \mu_F),
\end{align}
and $f_q(x, \mu_F)$, $q= c, \bar{c}$  the (anti-)charm distribution function in the proton, evaluated at the factorization scale $\mu_F$ which we identify in the following with the top quark mass. For numerical studies we use the leading order MMHT2014 PDF set \cite{Harland-Lang:2014zoa}.

\section{Model parameter space analysis for single top production}
\label{sec:num}

As our first approach to possible top production through neutral Higgs flavor violation processes, we perform a numerical analysis considering  $qq$ initial states, searching for same sign top pair production at LHC, with $0<\beta-\alpha<\frac{\pi}{2}$ as discussed in \cite{Gunion_2003},  and a relatively loose restriction on  $\tan \beta$, $\mathcal{O} (10^{-2})<\tan \beta<\mathcal{O} (10^{1})$ in order to observe the production behavior on this free parameters of the model, see Figure \ref{fig:exclude7}. At this point we want to state the following remark: too close to zero values on $\tan\beta$ would lead to an unwanted enhancement on u-quark type Higgs couplings which are proportional to $\cot\beta$ and, on the other hand, higher values on $\tan\beta$ would lead to an enhancement on d-quark type Higgs couplings, which in turn are proportional to $\tan\beta$. Our numerical analysis will show the behavior of the former case, having increasing values for $\sigma(pp \to tt, \bar{t}\bar{t})$ for low $\tan\beta$.

Furthermore, we calculate the scattering amplitude for the single top production within a flavor violation context, with a future perspective and contribution for proposed  lepton-proton colliders. Specifically, for the initial states $eq, \mu q$ and taking into account the neutral Higgs coupling in this extended model, we will get, $\mu \, t$ and $\tau \, t$ as part of the final states. Due to the form of the neutral Higgs boson couplings, these processes depend on all the masses of particles involved, we can see that for electron in the final state will greatly reduce the scattering amplitude.

The flavor violation coupling parameter, given in the matrix amplitude (\ref{Cs}), will be taking in general as $|\tilde{\chi}^{q,l}_{ij}|\sim \mathcal{O}(1)$, where $i,j$ stand for fermionic flavor\footnote{As can be seen in the analytical expression for the flavor violation cross sections, they are directly proportional to $|\tilde{\chi}^{q,l}_{ij}|^2$, any other values will change the cross section in this proportionality.}.  In calculating the cross section $ep\rightarrow l X_t X$ working in the context of the parton model, with the parton distribution function given in reference \cite{Pumplin:2002vw}, we observe that the charm-top transition is dominant over the up-top contribution, as is evident in Figure \ref{Fig.uvsc}. Hence, we can safely explore only the charm-top $\tilde{\chi}^u_{23}$ for hadronic part. 

In order to make our numerical calculations we will take the masses of the fermions involved in this work
 as follows: $m_e=0.511$ MeV, $m_{\mu}=105.66$ MeV, $m_{\tau}=1.777$ GeV, $M_p=0.938$ GeV, $m_u=2.16$ MeV, $m_c=1.27$ GeV and $m_t=172.69$ GeV. Finally, we take the $W{^\pm}$, $Z^0$ and $h_1^0$ as $m_{W^\pm}=80.44$ GeV, $m_{Z^0}=91.2$ and $m_{h_1^0}=0.125$ GeV \cite{ParticleDataGroup:2022pth}. 
  
\subsection{Constraints on charm-top transitions from LHC data}

In order to calculate the same sign pair top production via flavor violation at tree level, working in the context of 2HDM-III, 
we evaluate Eq.~\eqref{eq:Xsec:hh} for $\sqrt{s} = 7$~TeV, 
MMHT2014 leading order PDFs   \cite{Harland-Lang:2014zoa}, we  obtain the following numerical result,
\begin{align}
   \sum_{f=t,\bar{t}}    \sigma(pp \to ff)  & = C(\alpha, \beta) \cdot |\tilde{\chi}^u_{23}|^4 \cdot  9.37 \times 10^{-6}~\text{pb},
\end{align}
with  $C(\alpha, \beta)=\cos^4(\alpha-\beta)/\sin^4 (\beta)$.

From the upper limit given by ATLAS, we have $\sum_{f=t,\bar{t}}    \sigma(pp \to ff)  < 1.7$~pb, this allow us to establish a restriction on the model parameter's which is rather wide
\begin{align}
    C(\alpha, \beta) \cdot  |\tilde{\chi}^u_{23}|^4 < 1.81\times 10^{5}
\end{align}
For the current energy reached at LHC, $\sqrt{s} = 14$~TeV, the cross-sections for dedicated choices of the parameters $\alpha, \beta$, are given in Tab.\ref{LHC_ab}.

\begin{table}[ph]
\begin{center}
\caption{Cross-section for dedicated choices (some illustrating values along the range) of the model parameters $\alpha, \beta$ with $|\tilde{\chi}^u_{23}| =1$}
\begin{tabular}{|@{}ccc@{}|} 
\hline
$\beta-\alpha$ &$\tan\beta$&
$ \sum_{f=t,\bar{t}}\sigma(pp \to tt, \bar{t}\bar{t})$~(pb)\\ 
\hline\hline
$ \pi/3$ &$1/50$  &  $5.05$\\
\hline
$\pi/3$ & 1  &  $3.3 \times 10^{-6}$ \\
 \hline
 $\pi/3$ & 50 &  $8.1\times 10^{-7}$ \\
\hline
$\pi/4$ & $1/50$  &  $20$ \\
 \hline
$\pi/4$ & 1 &  $1.3\times 10^{-5}$ \\
\hline
$\pi/4$ & 50   &  $3.2\times 10^{-6}$ \\ 
\hline
\end{tabular}
\label{LHC_ab} 
\end{center}
\end{table}

\begin{figure}[htbp]
     \centering
\includegraphics[width=0.40\textwidth]{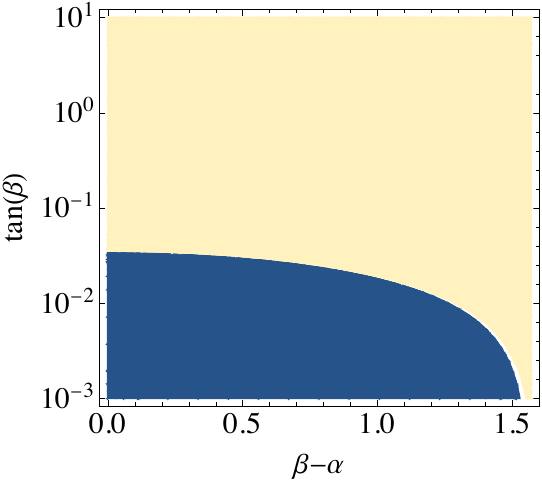}
\includegraphics[width=0.46\textwidth]{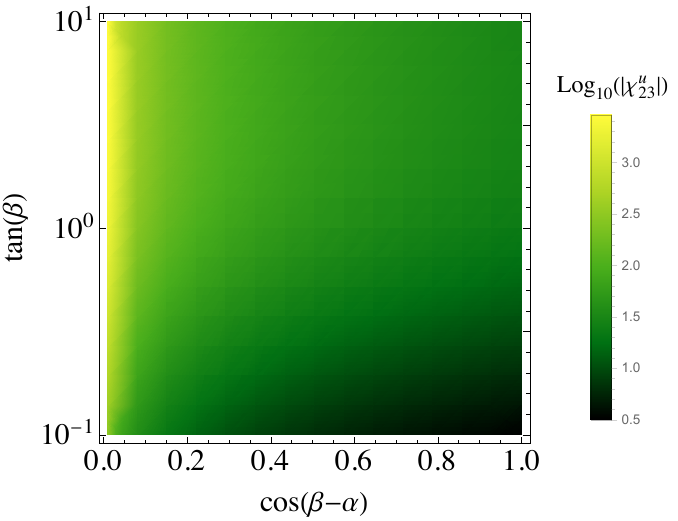}
     \caption{Same sign double top production considering the experimental bound for $\sigma(pp \to tt, \bar{t}\bar{t})<1.7$~pb. ({\it Left}) Setting $|\chi_{23}^u|$ = 1, in blue the range of parameters excluded by current $7$~TeV LHC data.  ({\it Right}) Excluded values of $|\chi_{23}^u|$ depending on given values of $\tan \beta$ and $\cos(\beta - \alpha)$. }
     \label{fig:exclude7}
 \end{figure}

We can see in  this Table \ref{LHC_ab} that as $\beta-\alpha$ is closer to $\pi/2$ the cross-section diminishes, recovering the SM prediction in the limit $\beta-\alpha \to \pi/2$, while the cross-section increases for low $\tan\beta <1$. Fig.~\ref{fig:exclude7} illustrates the  parameter space region excluded by $7$~TeV LHC data, where we observe that $\tan\beta \lesssim 5\times 10^{-2}$ gets strongly suppressed from experimental data for the case of $\beta-\alpha\lesssim \pi/2$, seen as a excluded blue part in the Figure \ref{fig:exclude7}. 
 For $\beta-\alpha \to \pi/2$   the cross-section diminishes and we recover the SM prediction, while the cross-section is enhanced for $\tan\beta <1$. Fig.~\ref{fig:exclude7} illustrates the  parameter space region excluded by $7$~TeV LHC data. Fig.~\ref{fig:exclude7}, left considers the special case of  $|\chi^u_{23}|=1$. 

Fig.~\ref{fig:exclude7}, right finally explores the bound on $|\chi^u_{23}|$, if $\alpha, \beta$ are taken as free parameters. We find that higher possible values for quark flavour violation parameter of the model $|\chi_{23}^u|\gtrsim 10^3$ are allowed only if $\cos(\beta - \alpha)\approx 0$ and  $\tan \beta \gtrsim 10^{-1}$. On the other hand, $\tilde{\chi}^u_{23}\lesssim 1$ then,  $\tan \beta \lesssim 10^{-1}$ and $\cos(\beta - \alpha)\gtrsim 0.2$.

\subsection{FCNC single top production through $e q\to l+X_t +X$ }

The next process we analyze is given as $ep\to l X_t X $, where $l$ is a charged lepton, $X_t$ corresponds to a jet coming from top production and $X$ could be anything.
From the analytical expression given in previous section we are in a position to discuss the parameter space for electron-proton collisions related to the production of a single top through FCNC. 
Also we explore the process for a potential muon-proton collider \cite{https://doi.org/10.48550/arxiv.2203.06258} which will be decisive for this FCNC.

Observed that, as we said before, the processes are via Higgs boson exchange, the amplitude of the processes depend on the masses of the external particles involved, as shown in eq. (\ref{Mampli}) and the parameters of the model eq. (\ref{Cs}). 
Three processes considered here are dependent through the different parameters $\tilde{\chi}^l_{ij}$ and the masses of the leptons involved. 
Each cross section, $\sigma(l_i q_{k}\rightarrow l'_j t X)$ with  ($i\neq j$) have one leptonic parameter $\tilde{\chi}^{l}_{ij}$ and in principle, two quark contributions $\tilde{\chi}^{u}_{k3}$,  with $k=1,2$. Although there is no relation between processes with different leptons, the parameters are independent from one another, we will show in our numerical analysis, the charm quark contribution $|\tilde{\chi}^{u}_{23}|^2$ dominates, then we can factorize the parameters, then
we can express these relations as follows:
\begin{eqnarray}
    \sigma (ep \to \tau q X) \approxeq \frac{m_\tau}{m_\mu}\frac{|\tilde{\chi}^l_{13}|}{|\tilde{\chi}^l_{12}|}\sigma (ep \to \mu q X) \nonumber \\
    \sigma (\mu p \to \tau q X)  \approxeq  \frac{m_\tau}{m_e}\frac{|\tilde{\chi}^l_{23}|}{|\tilde{\chi}^l_{12}|}\sigma(ep \to \mu q X);
    \label{sigmaRel}
\end{eqnarray}
The differences could come from the kinematic relations, nevertheless in this paper, for the numerical calculation purpose, we consider in the kinematics, all fermion massless except for the quark top. Hence, in this approximation for the kinematics we reach the equality on Eq. (\ref{sigmaRel}) when all lepton masses are taken equal to zero, and only in the dynamics we consider the masses involved different from zero.

For the numerical calculation of the sub-processes $l_iq\rightarrow l_j t $ cross section, we consider first a range for $\cos(\beta-\alpha)\in [0,1]$ and $\tan \beta \in [1,50]$ (discussions about the THDM and parameter space  can be found in \cite{Gunion_2003,Branco_2012}), for energies in the center-of-mass running from $\sqrt{s} \in [1.3,50]$ TeV.

The results we obtained for processes with lepton flavor violation are given in general  as
$$\sigma(ep\rightarrow \mu X_t X)=\mathcal{O}(10^{-5})|\tilde{\chi}^{l}_{12}|^2|\tilde{\chi}^{u}_{23}|^2$$
The other processes of this type will be obtained by Eq. (\ref{sigmaRel}).

\begin{figure}[h!]
\centering
\includegraphics[width=0.8\textwidth]{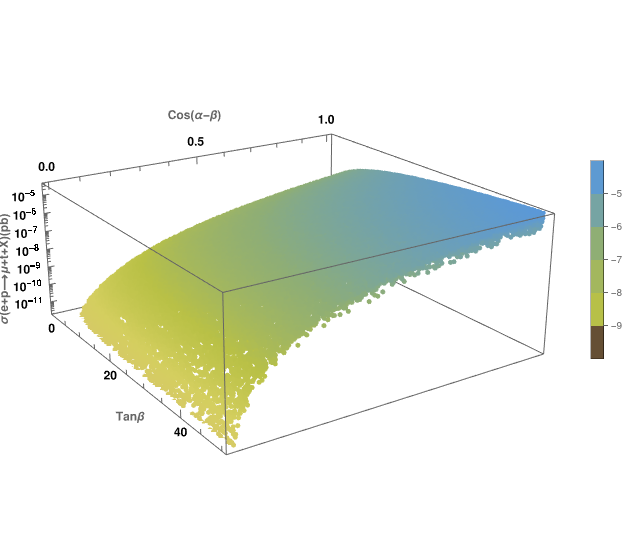}
\caption{The total cross section to the process $e p\rightarrow \mu+ X_t+X$, with center of mass energies $\sqrt{s}\in[1.3, 50]$ TeV. Is is shown its dependence on the wide range of free parameters $\tan\beta \in [1,50]$, $\cos(\beta-\alpha)\in [0,1]$ and $|\tilde{\chi}_{12}^l|=|\tilde{\chi}_{ij}^u|=1$. }
\label{Fig.3Dep-mut}
\end{figure}

As seen in Fig. \ref{Fig.3Dep-mut}, the cross sections  will increase to its maximal value for $\cos(\beta-\alpha)\approx 1$ and high values of $\tan\beta\sim 50$. In this case, the order of magnitude of the parameters are taken as $\vert\tilde{\chi}_{ij}^{u,d,l}\vert^{2}=1$, as is suggested in the article of 2HDM-III\cite{Diaz-Cruz:2004hrt}.

\begin{figure}
\centering
\includegraphics[width=0.8\textwidth]{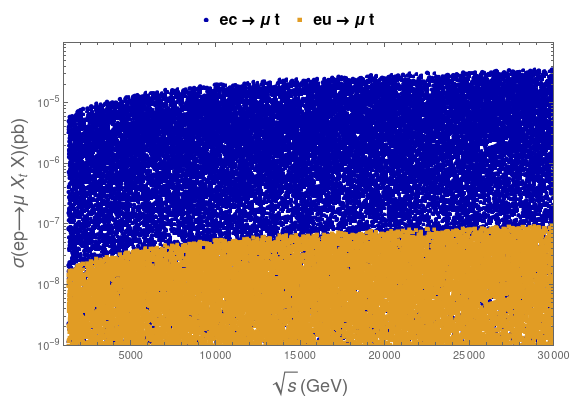}
\caption {The contribution from each quark, $u$(green), $c$(blue), to the total cross section  to the process $e q\rightarrow \mu t$, with $q$ stands for $u,c$. Considering the values of the parameters as $\tan\beta \in [1,50]$, $\cos(\beta-\alpha)\in [0,1]$ and $|\tilde{\chi}_{12}^l|=|\tilde{\chi}_{13}^u|=|\tilde{\chi}_{23}^u|=1$.}
\label{Fig.uvsc}
\end{figure}

As we already said, we see from our results in Fig. \ref{Fig.uvsc} that the parton contribution from u-quark is small compared with the dominating c-quark contribution, so we can safely explore the single top production process exclusively from the charm quark contribution.

\begin{figure}[h!]
\centering
\includegraphics[width=0.8\textwidth]{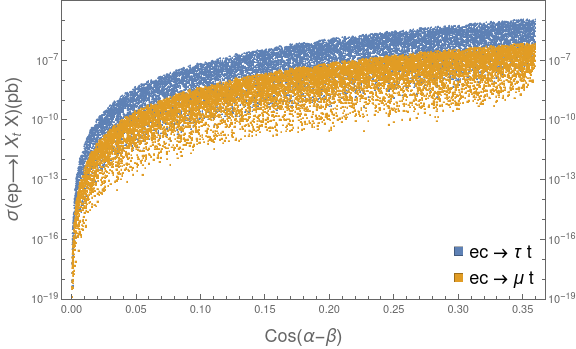}
\caption{The total cross section for $e p$ t-channel FCNC top production via DIS dependence with $\cos(\alpha-\beta)$. Comparing two leptonic processes: $e c\rightarrow \tau (\mu)+t $, {\it blue (yellow)}. With a scan for the center of mass energy as $\sqrt{s}\in[1.3, 50]$ TeV, applying the restriction on top FV neutral Higgs boson decay, (\ref{tchBound}). }
\label{ep2lep-t-cna}
\end{figure}
Considering the experimental bounds over flavor violations process as top decay $\Gamma(t\to c h)$, the parameter would be restricted as is reported in \cite{arhrib2016two},
\begin{eqnarray}
 \Gamma(t \to c h)&=&\frac{1}{32\pi m_t^2}\left( \frac{\cos(\beta -\alpha)\tilde{\chi}^u_{23}}{\sin \beta} \right)^2\left(
 (m_c+m_t)^2-m_h^2\right) \nonumber \\
 &\times &\sqrt{(m_t^2-(m_h-m_c)^2)(m_t^2-(m_h+m_c)^2)}
\end{eqnarray}

At the LHC, ATLAS and CMS have searched for top FCNCs and they set a limit on the flavor violating branching fraction $Br(t \to ch) < 0.82\%$ for ATLAS \cite{TheATLAScollaboration:2013nbo} and $Br(t \to ch) < 0.56 \%$ for CMS \cite{CMS:2014qxa}. We will use the bound given for CMS, as is the more stringent because they take into account more processes in the search. 

Then, using the CMS results to bound the free FV parameters of the model we get
\begin{eqnarray}
 \left( \frac{\cos(\beta -\alpha)\chi^u_{23}}{\sin \beta} \right)< 0.36 & \text{  for  } & Br(t\to c h)< 5.6\times 10^{-3}
 \label{tchBound}
\end{eqnarray}
In general, the experimental bound for a flavor violation decay $y^{Exp}_{ij}$ will generate a relation with the free parameters $\alpha, \beta $ and $\tilde{\chi}^f_{ij}$ as follows
\begin{eqnarray}
    \left|\frac{|\tilde{\chi}_{ij}|^2 \cos \alpha}{y^{Exp}_{ij}-\sin\alpha|\tilde{\chi}_{ij}|^2}\right|<\tan \beta
\end{eqnarray}

From figure \ref{Fig.uvsc} we can see that the dependence on the center on mass energy $\sqrt{s}$ is not significant, the relevance will come from improvement on luminescence.

Our results for $e p\rightarrow \tau (\mu)+t $ are shown in Fig.~\ref{Fig.uvsc} and Fig.~\ref{ep2lep-t-cna}. 
Fig.~\ref{Fig.uvsc} shows  electron and up quark ($eu$) and electron and charm quark ($ec$) at the initial state; note that convolutions with the corresponding parton distribution functions have been taken into account. 

In Fig.~\ref{ep2lep-t-cna} we present the comparison of the production of single top with muon, $e p \to \mu + X_t + X $ and the single top production accompanied with a tau, in the process $ e p \to  \tau + X_t +X $, at $\sqrt{s}=1.3 - 50$~TeV; considering the restriction on the parameters given in Eq. (\ref{tchBound}). 
As expected according to Eq.(\ref{sigmaRel}), the difference is given by $m_\tau/m_\mu$.

\subsection{FCNC single top production through $\mu p\to l+X_t +X$ }
For completeness, in this subsection we include the analysis of single top production in a possible $\mu p$ scattering processes. The muon collider has been studied in terms of the energies and the kinematics achieved \cite{acosta2022muon}, then we explore possible exotic flavor processes.
We find that in this kind of colliders, the flavor violation single top production, through DIS scattering will improve the probability to detect this exotic signal.
Moreover, as we explored the analytical expressions in section 3, the lepton-lepton-Higgs coupling has the special structure that includes SM and also 2HDM contributions, which in turn could have opposite signs, and then enhance or diminish the scattering amplitude having no direct proportionality with the masses and the $|\tilde{\chi}^l_{ii}|^2$ values, this is why detailed numerical exploration is more relevant in this case.

\begin{figure}[hbt!]
\centering
\includegraphics[width=0.8\textwidth]{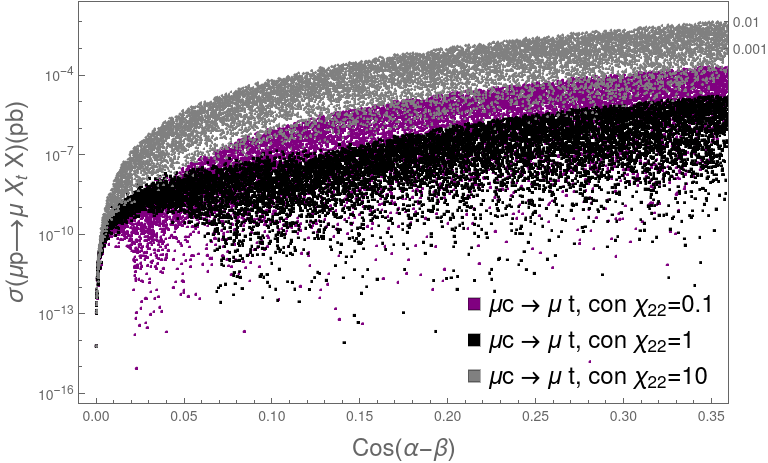}
\caption{\normalsize{Cross section for t-channel FCNC  top production via DIS $\mu c \to \mu t $ comparing different orders of FV parameter: $\tilde{\chi}^l_{22}=0.1,1,10$ in {\it purple, black} and {\it gray}, respectively. Within a range for the center of mass energies $\sqrt{s}\in[1.3, 50]$ TeV, considering the $t \to c h$ restriction (\ref{tchBound}).}}
\label{fig:resultsmu2p}
\end{figure}
\begin{figure}[hbt!]
\centering
\includegraphics[width=0.8\textwidth]{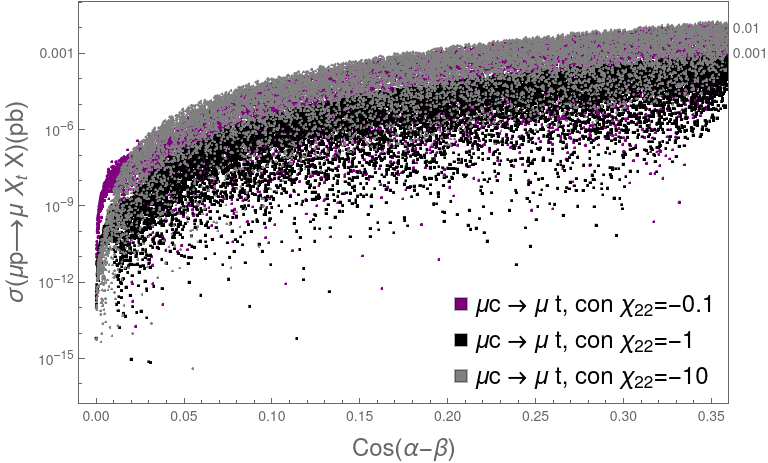}
\caption{\normalsize{Cross section for t-channel FCNC  top production via DIS $\mu c \to \mu t $ comparing different orders but negative value of FV parameter: $\tilde{\chi}^l_{22}=-0.1,-1,-10$ in {\it purple, black} and {\it gray}, respectively. Within a range for the center of mass energies $\sqrt{s}\in[1.3, 50]$ TeV, considering the $t \to c h$ restriction, (\ref{tchBound}).}}
\label{fig:resultsmu2n}
\end{figure}

As we discuss in section 3, the flavor violation parameter could be complex, for purpose of illustration we take extreme values for the complex  phases, positive and negative, then we would have possibilities of constructive or destructive interference with the SM term, see Eq. (\ref{Cs3}). These results are shown in Fig. \ref{fig:resultsmu2p}, for positive values, and in Fig. \ref{fig:resultsmu2n} for negative values. An aspect that is worth to notice in the positive case is that for $\tilde{\chi}^l_{22}=0.1$ we would have grater values for the cross section than those for $\tilde{\chi}^l_{22}=1$, implying that for the last case, the destructive interference is greater. For a high value of the parameter, $\tilde{\chi}^l_{22}=10$ we reach 0.01 pb for the single top production cross section, and the 2HDM contribution being the dominant one. 
For negative values of the parameter, Fig. \ref{fig:resultsmu2n}, we observe a general constructive interference with the SM contribution.

\begin{figure}[hbt!]
\centering
\includegraphics[width=.8\textwidth]{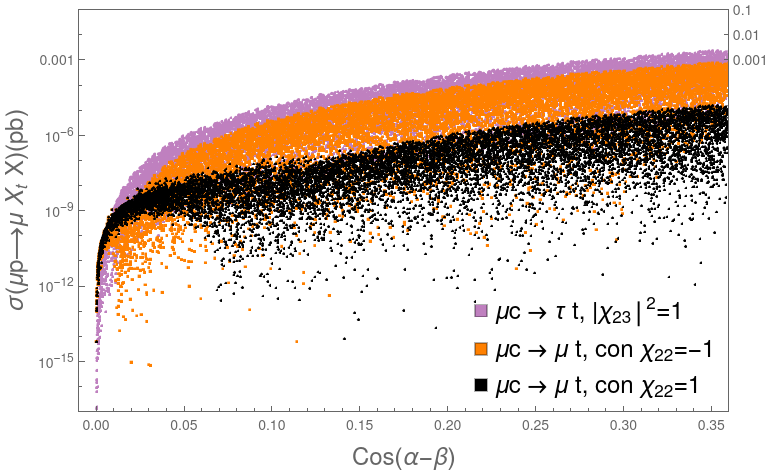}
\caption{Cross section for t-channel FCNC  top production via DIS comparing $\mu c \to \mu  t $, setting $\tilde{\chi}^q_{23}= 1$ and $\tilde{\chi}^l_{22}= \pm 1)$, positive and negative, {\it black} and {\it orange}, respectively. For $\mu c \to \tau  t $, we take $|\tilde{\chi}^l_{23}|^2 = 1)$ in {\it purple}; within a range for the center of mass energies $\sqrt{s}\in[1.3, 50]$ TeV, considering the $t \to c h$ restriction, (\ref{tchBound}).}
\label{fig:resultsmu2mutau}
\end{figure}

In Fig. \ref{fig:resultsmu2mutau}, as is expected, the single top production with muon-tau flavor violation process increase the cross section value. It is also seen here, that for negative values of the parameter, the interference is constructive, compared with the positive one of the same order.

\section{Conclusions}
Rare top processes could be a key analysis to establish the model structure of beyond the Standard Model physics in terms of the yukawa structure of the Higgs couplings and degrees of freedom which allow for flavor violating. 
In this paper we discussed the possibility to observe effects of FCNC through flavor violating neutral Higgs boson accompanied by either same sign top pair production for $pp$ collisions, or single top production at lepton-proton colliders. 

While we found that cross-sections are small, $\sigma (f q \to f' t)\lesssim \mathcal{O}(10^{-1})pb$, and its observation would be challenging. On the other hand, also means that the consideration of BSM FV couplings in the Higgs sector at tree level could not be ruled out from current experiments, a need for future colliders lepton-proton results that would be more decisive.
We performed the complete analytical calculation for exotic flavor violation process in a 2HDM in proton-proton and lepton-proton collisions at LO.
In general, FCNC
interactions,  where neutral Higgs boson and top quark are involved, are an excellent source to seek beyond Standard Model signals.
The analytic structure for each of these processes were generated through the corresponding sub-processes $lq\to l't$ and  $qq\to tt$, with $q=u,c$. Considering that top quark does not contribute to the sea quark, due to its heavy mass,  we use PDFs based in a five light quark scheme.
In addition, according to 
\cite{4} those kind of processes have not been fully reviewed so far, then, continuing the exploration for different scenarios would be relevant to establish BSM possibilities. 
In order to search for clean signals of new physics, which would directly confirm or discard this FV process through Higgs neutral currents, we determined a specific non-SM analogous tree level process through t-channel in Deep Inelastic Scattering  that could be tested at current and future colliders.

For the analysis of the FV process, we first proceed calculating same sign pair top production at a proton-proton collider, $p+p\to t+t+X$ which establish a first parameter restriction from LHC, given by both 7TeV and 14TeV LHC data for 2HDM parameter space. We got a restriction for 2HDM parameters, $\tan\beta \gtrapprox 5\times 10^{-2}$ with $\beta-\alpha \lesssim \pi/3$.
We found that the upper limit is very loose as allows even the FV coupling parameter to be up to $\tilde{\chi}_{ij} \sim \mathcal{O}(10)$. Then, we focus on single top production processes, for electron-proton collider $l+p\to l'+t+X$ and use the bound given by HERA, although its energy reach is low. The future EIC collider will not have enough energy to produce a top quark as it would have variable center of mass energies from $\sim 20 \to \sim 100$ GeV, up gradable to $\sim 140$ GeV, \cite{Accardi:2012qut,AbdulKhalek:2021gbh}.

It is clear from the Feynman diagrams of the sub-processes that there is no counterpart within the SM, these are necessarily BSM processes.
From the structure of the analytical scattering amplitude and also corroborating with a numerical analysis which includes the PDFs, we found that the c-quark contribution dominates the process over the u-quark, because the masses are relevant directly on couplings than the valence nature of the quark.  

We also found that the process will not be sensible to the energy scale. There is no difference for the calculated cross section for an energy range from 1.3 TeV to 50 TeV.  Rather the luminosity will be the most important factor that may enhance the possibility of measure this exotic processes.

We extended our calculation to illustrate the enhancement in a muon-proton collider, because the particle masses are involved in the couplings. The $\mu q\to \mu t$ process most include the SM-type muon vertex and the FV contribution from 2HDM. The analytical structure differs from the doble flavor violation in lepton and quark couplings. We found in this case, that as the  complex structure of $\tilde{\chi}_{ij}$ is relevant, we explore negative sign possibilities obtaining an enhancement due to the fact that the SM-type coupling is negative. \\

 We showed the model parameter space dependence  for the  scattering cross section. Enhancing the cross section $\sigma(pp\to t t (\bar{t}\bar{t})+X)$ up to $\sim \mathcal{O}(10) pb$, for low $\tan\beta\sim \mathcal{O}(10^{-2})$ and for  $\cos(\alpha-\beta)\sim 1$. And  $\sigma(ep\to \mu t+X)\sim 10^{-5}pb$, for large $\tan\beta\approx 50$ and $\cos(\alpha-\beta)\approx 1$. The cross section is enhanced by two orders of magnitude when we consider muon-proton collider and tau involved.
The experimental bound on $\Gamma(t\to c h)$ from ATLAS and CMS restricts the values of the 2HDM FV parameters. We used the most restrictive experimental value which coming from CMS, restricting the values of $\cos(\beta-\alpha)\lesssim 0.35 $ considering $|\chi_{ij}|^2\approx 1$.
Even if the process is not measured, due to its small value, bounds on FV parameter of the model, $\tilde{\chi}_{ff'}$, are achieved as the analytical expression is directly proportional to this parameter.



\section*{Acknowledgements}
We want to thank Cristian Baldenegro for his useful discussion on experimental results. M. G\'omez-Bock acknowledge the support by UNAM under IN109321 PAPIIT. W. Gonzalez gratefully acknowledges the scholarship from CONACyT which allowed development of the present work and also thankful to Dr. A. Rosado for for financial support at the project. Support by  Consejo Nacional de Ciencia y Tecnolog\'ia grant number
A1 S-43940 (CONACYT-SEP Ciencias B\'asicas) as well as funds assigned
by the Decanato de Investigaci\'on y Posgrado UDLAP are gratefully
acknowledged. 
MGB dedicate this work to Sandra Rodriguez Pe\~na, for her great animosity towards this work, and essentially towards life.

\appendix


\bibliographystyle{unsrt}
\bibliography{referencestop}


\end{document}